# Additive-Induced Stabilization of the Energetic Landscape of PM6:Y12 Organic Solar Cells

*Bekcy Joseph,[1,2] Shivam Singh,[1,2] Nathaniel P. Gallop,[1,2] Fabian Eller,[3] Alexander Ehm,[4] Julius Brunner,[1,2] Dietrich R. T. Zahn,[4] Eva Herzig,[3] Boris Rivkin,[1,2] Yana Vaynzof\* [1,2]*

[1]Chair for Emerging Electronic Technologies, Technical University of Dresden, Nöthnitzer Str. 61, 01187 Dresden, Germany.
[2]Leibniz-Institute for Solid State and Materials Research Dresden, Helmholtzstraße 20, 01069 Dresden, Germany.
[3]Dynamics and Structure Formation – Herzig Group, University of Bayreuth, 95440 Bayreuth, Germany
[4]Semiconductor Physics and Research Center for Materials Architectures and Integration of Nanomembranes (MAIN), TU Chemnitz, 09107 Chemnitz, Germany

y.vaynzof@ifw-dresden.de

**Abstract**

Solvent additive engineering is a common strategy in organic photovoltaic (OPV) fabrication to improve film morphology and enhance device performance by controlling phase-separation kinetics and crystallinity. However, its effect on photostability, particularly with respect to the evolution of the energetic landscape under operational stress, remains unclear. This study investigates the impact of the additive 1-chloronaphthalene (1-CN) on the evolution of the device's energetic landscape in PM6:Y12 bulk heterojunction organic solar cells upon photoaging. Ultraviolet photoemission spectroscopy combined with argon gas cluster ion beam depth profiling is employed to probe the depth-resolved evolution of donor (PM6) and acceptor (Y12) energy levels before and after photodegradation. Our findings show that in additive-free devices, photodegradation leads to a significant 200 meV downward shift in the PM6 highest occupied molecular orbital (HOMO) level, reducing the donor-acceptor HOMO offset and impairing the driving force for hole transfer. As a consequence, the device experiences substantial efficiency loss. On the other hand, the incorporation of 1-CN effectively stabilizes the PM6 HOMO level, preserving adequate driving force for efficient exciton dissociation. Advanced X-ray diffraction characterization reveals more pronounced nanostructural degradation in blends without 1-CN than those with 1-CN upon photoaging. Collectively, these findings identify PM6 as the primary degradation pathway in PM6:Y12 blends and demonstrate that 1-CN enhances device stability by stabilizing PM6 energetics and preserving the nanostructural integrity upon photoaging.

## 1. Introduction

Organic photovoltaic (OPV) devices have emerged as a promising renewable energy technology, achieving power conversion efficiencies (PCE) exceeding 20% in recent years[1–5]. The development of non-fullerene acceptor (NFA)-based solar cells has enabled the transition from fullerene-based systems to more efficient, tunable architectures with reduced voltage losses and broad absorption in the near-infrared region[6–9]. Despite their impressive efficiency, potential for low cost, and flexible manufacturing, the commercialization of OPV is still hindered by its inability to achieve simultaneous high efficiency and long-term stability [10–13]. Therefore, comprehending the degradation mechanisms in OPV is crucial for advancing the development of commercially viable OPV devices.

Degradation in OPV devices is influenced by multiple interconnected factors, including chemical degradation of organic materials, interfacial instability, and morphological changes in the active layer[14–16]. Among these pathways, the instability of the nanoscale bulk heterojunction (BHJ) morphology represents one of the most critical challenges. Several strategies have been developed to improve morphological stability, including molecular design of donors[17–19] and acceptors[20,21], ternary blends[22–25], and solvent engineering[26,27]. Among these approaches, solvent additive engineering has emerged as a particularly effective and widely adopted strategy. The incorporation of solvent additives into the active layer has emerged as an effective strategy to regulate solvent evaporation kinetics during the transition from solution to film, thereby enabling precise control over film morphology and consequent enhancement of photovoltaic performance[28–30]. The morphology optimization induced by additive engineering leads to reduced charge recombination, enhanced exciton dissociation efficiency, and more balanced charge transport compared to pristine devices[31,32]. Among the additives,1-chloronaphthlene (1-CN) is one of the widely used additives, due to its high boiling point, which slows the solvent-evaporation process and selectively modulates donor (D)-acceptor (A) solubility to optimize BHJ morphology[33–36]. Numerous studies have demonstrated that the addition of solvent additives like 1-CN improves device performance through higher crystallinity, greater donor-acceptor interfacial area, and more refined phase separation[34,37–40]. However, while the performance benefits of additive engineering are well established, its influence on the long-term operational stability of OPV devices, particularly regarding photodegradation mechanisms and the evolution of energetic properties during photoaging, remains unclear.

The donor-acceptor interface is a critical factor in optimizing device performance, as it governs excitation dissociation and subsequent free charge generation[41–43], ultimately affecting device efficiency and operational stability[44,45]. In BHJs incorporating low-bandgap NFAs, both the electron donor and acceptor materials contribute to light harvesting, giving rise to two distinct charge generation pathways: electron transfer from the donor to the acceptor, driven by the lowest unoccupied molecular orbital (LUMO) offset, and hole transfer from the acceptor to the donor, driven by the highest occupied molecular orbital (HOMO) offset. Though the electron transfer from donor to acceptor has been studied as the primary charge generation pathway in BHJ, for low bandgap NFAs, hole transfer from the low bandgap acceptor to the donor is the key step for free charge generation, and the efficiency of this process is determined by the HOMO offset between donor and acceptor[46–48]. Yang et al. demonstrated that in PM6-based non-fullerene OPV, a sufficient energy level offset between donor and acceptor is necessary to provide adequate driving force for efficient exciton dissociation, and when the HOMO offset falls below a critical threshold, the exciton dissociation process is severely hindered, thereby deteriorating device performance[49]. Similarly, Karuthedath et al. established that for low-bandgap NFA-based OPVs, charge generation depends primarily on hole transfer from acceptor to donor, and that a sizeable bulk HOMO offset is essential for designing efficient low-bandgap NFA-based OPVs.[50] While these studies have conclusively demonstrated that a threshold HOMO offset is necessary to sustain efficient charge generation, the evolution of these critical energetic offsets during device aging and photodegradation remains unexplored.

Conventionally, energy level offsets between individual donor and acceptor materials are measured separately using techniques such as cyclic voltammetry (CV), Kelvin probe spectroscopy, or ultraviolet photoemission spectroscopy (UPS), and these values are then assumed to represent the energetics within the BHJ blend. However, interfacial effects can occur due to the strong blend of donor and acceptor, which can influence the electronic alignment at the interface and the formation of interfacial dipoles through charge transfer, thereby altering the effective energy level offsets from those predicted from separate measurements[44]. Also, surface-sensitive measurements like UPS provide information only about the surface, and it is inadequate to assume that the bulk of the active layer behaves similarly to the surface, as vertical compositional gradients, surface enrichment of one component, and differential exposure to ambient conditions can all lead to significant differences between surface and bulk energetics[51]. Bilayer systems exhibit interfacial dipoles at the donor-acceptor interface that cannot be predicted or explained solely from surface energy level measurements[41,44]. These limitations necessitate the development of techniques capable of probing the actual energetics throughout the depth of the BHJ active layer.

To address these limitations and probe the realistic energetics throughout the bulk active layer, we employ UPS combined with argon gas cluster ion beam depth profiling (UPS-GCIB-DP)[51]. In this technique, argon gas cluster ions are used to progressively and gently remove material from the surface, enabling the acquisition of UPS spectra at defined depth intervals across the full thickness of the BHJ without significantly damaging the organic semiconductor or altering its electronic structure. It helps to acquire the energetics at the D-A interface, not from traditional surface measured energy levels, but the realistic energetics at BHJ. The technique can be applied at any stage of device operation, making it a powerful tool for investigating the evolution of electronic profiles during degradation. Here, we employ the technique to understand the evolution of energetics in BHJ upon photodegradation. Under continuous illumination, the device can experience shifts in its energetic properties, namely changes in the HOMO levels of polymer-low-bandgap NFAs, which subsequently compromise charge extraction efficiency and overall device performance.

In this work, we systematically investigate how photodegradation impacts the energetics of PM6:Y12 BHJ and examine the influence of 1-CN additive on the energetic stability and operational performance of these BHJ OPV devices. The degradation studies are conducted on conventional organic solar cells with the structure ITO/PEDOT: PSS/PM6:Y12/PDINN/Ag. 0.5% v/v 1-CN additive is incorporated into the active layer to investigate its effect on the stability and performance of PM6:Y12 solar cells. The chemical structure of the polymer PM6, non-fullerene acceptor Y12, and the additive 1-CN are depicted in Figure 1a. As large-scale roll-to-roll fabrication of OPV modules is typically carried out under ambient conditions, which permits oxygen and water into the bulk layer, the films in this study were aged under continuous 1 sun illumination (AM 1.5G, 100 mW cm$^{-2}$) in ambient air for 15 h. Freshly prepared films and devices are labelled as fresh, and aged films are labelled as aged throughout the study. By integrating depth-resolved UPS energetics, GIWAXS nanostructure analysis, transient absorption charge dynamics, and PV performances, we elucidate how additive engineering influences the energetic landscape and degradation pathways in PM6:Y12 BHJ OPVs.

## 2. Results and Discussions

### 2.1 Optical Properties of PM6 and Y12

To understand how photodegradation affects the optical properties of the donor and acceptor materials, we first examined the absorption spectra of fresh and aged neat PM6 and Y12 films, as well as PM6:Y12 blend films with and without 1-CN (Figure 1b-e). The absorption spectra of fresh PM6 and Y12 neat films exhibit characteristic absorption maxima at 615 nm and 837 nm, respectively. After 15 hours of continuous light soaking under 1 sun illumination in ambient air, the PM6 absorption peak at 615 nm decreases significantly in magnitude. After 15 hours of continuous light soaking under 1 sun illumination in ambient air, the PM6 absorption peak at 615 nm decreases significantly in magnitude. Also, the vibronic shoulder structure of this feature is completely lost in the neat PM6 film (Figure 1b), indicating a reduction in conjugation length and structural order along the polymer chain [52]. In contrast, the Y12 acceptor does not show any significant change in the absorption peak in the aged film (Figure 1c), demonstrating the remarkable photochemical stability of this non-fullerene acceptor. A similar trend is observed in the PM6:Y12 blend films (Figure 1d), where the vibronic shoulder of PM6 is absent, but it remains intact for Y12. However, the reduced intensity is attributed to photobleaching. The pronounced spectral instability observed specifically in PM6 absorption features provides strong initial evidence that the polymer donor is the component most severely affected by photoaging, while the non-fullerene acceptor Y12 remains largely stable.

When 1-CN additive is incorporated into the PM6:Y12 blend, optical degradation is substantially mitigated. By comparing Figures 1d and 1e, it is evident that the addition of 1-CN to the PM6:Y12 blend has largely preserved the overall photon absorption upon aging compared to the blends without 1-CN. Notably, the addition of 1-CN to the active layer results in noticeable sharpening and increased intensity of the donor and acceptor peaks, as well as the formation of a distinct shoulder near 747 nm, indicating enhanced aggregate formation or favourable intermolecular interactions at the donor–acceptor interface[53]. These spectral changes collectively suggest that additive engineering with 1-CN substantially improves molecular ordering and packing in both the fresh and aged blend films.

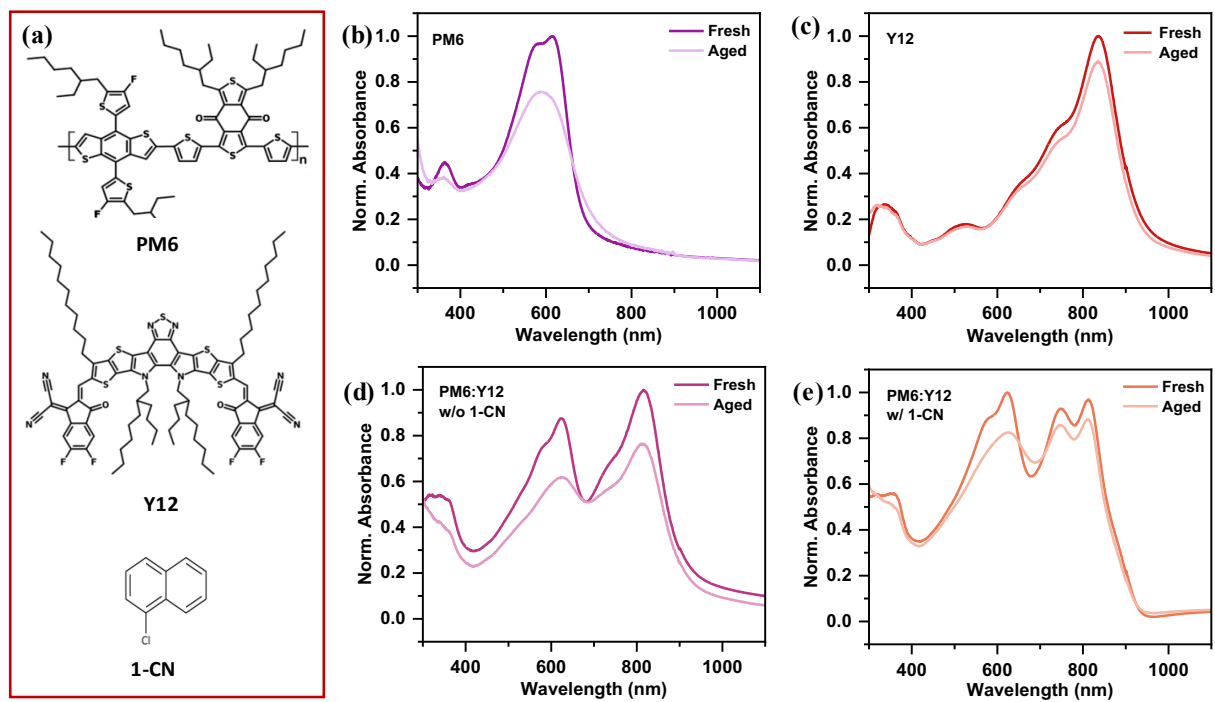

**Figure 1.** (a) Chemical structure of the PM6, Y12, and 1-CN. Absorption spectra of fresh and aged (a) PM6, (b) Y12, (c) PM6:Y12 without 1-CN, (d) PM6:Y12 with 1-CN.

To gain further insight into energetic disorder and sub-bandgap states in PM6, we performed photothermal deflection spectroscopy (PDS) measurements (Figure S1). The Urbach energy, extracted from the exponential tail of the sub-bandgap absorption, provides a quantitative measure of energetic disorder in the material. For neat PM6 films, the Urbach energy increases significantly after degradation, from 95.8 meV in a fresh film to 118.4 meV in an aged film, indicating substantially increased energetic disorder in the donor material after photoaging, consistent with the formation of trap states and the loss of structural order reported in the literature [52]. Interestingly, PM6:Y12 blend films show a much smaller change in Urbach energy, with the fresh film exhibiting 20.9 meV and the aged film exhibiting 22.6 meV. This minimal change in blend films can be attributed to the dominant contribution from Y12, which has a low-energy absorption tail that largely determines the absorption, thereby showing a lower energetic disorder. Nevertheless, the absorption and PDS spectra collectively confirm that PM6 domains are primarily affected by aging.

## 2.2 Evolution of Energetic Landscape of PM6:Y12 upon Photodegradation

Individual UPS spectra for neat PM6 and Y12 materials are presented in Figure S2. As a surface-sensitive technique, conventional UPS provides information primarily within the top 1–2 nm of the sample surface. Thus, to obtain insights into the vertical energetic profile across the entire bulk heterojunction (~100 nm thickness), UPS depth profiling was employed. This method uses argon gas cluster ion beam etching to progressively remove material, enabling the acquisition of UPS spectra at defined depth intervals. The resultant valence band spectra of BHJ containing both donor and acceptor are fitted by a linear combination of the valence band spectra of individual materials. This quantitative deconvolution helps to extract the HOMO levels for the donor and acceptor as a function of depth. The secondary electron cut-off (SECO) and the low-binding-energy edge near the Fermi level (HOMO onset) are used to extract the vacuum level and the HOMO levels of the donor and acceptor, respectively. LUMO levels were calculated by subtracting the optical bandgap from the HOMO level of each material. Electron affinities of donor and acceptor were assessed using Low-Energy Inverse Photoemission Spectroscopy (LEIPES), as depicted in Figure S3.

Three-dimensional maps of the energetic landscape near the HOMO region, derived from UPS depth profiles for fresh and aged PM6:Y12 blends with 1-CN, are shown in Figure 2a and 2 b. The full UPS spectra, including both the HOMO and SECO regions, are provided in Figure S4. It shows that both fresh and aged PM6:Y12 blends with 1-CN exhibit uniform energy levels, indicating electronic homogeneity across the layer thickness. The PM6:Y12 blends also preserve the electronic uniformity of the BHJ layer (Figure S4).

The Vertical gradient of energy levels in the PM6:Y12 active layer, determined from UPS depth profiling, is presented in Figures 2c and 2d. A slight upward bending (toward the vacuum level) of the energy levels is observed at the film surface, in both fresh and aged blend films. This surface band bending can be attributed to the films' exposure to moisture and atmospheric oxygen during fabrication. However, the HOMO onsets and Fermi level remain uniform throughout the bulk of the active layer. In PM6:Y12 blends, photodegradation induces a critical 200 meV downward shift of the PM6 donor HOMO level from −5.3 eV to −5.5 eV (Figure 2c). The instrumental experimental error is approximately 0.1 eV. In contrast, the Y12 acceptor demonstrates exceptional energetic stability, with its HOMO level remaining essentially invariant at −5.50 eV before and after photodegradation. The stability of Y12 energetics confirms that the energetics of the acceptor are not significantly affected by the aging, and the PM6 donor is directly affected by degradation.

Here, the LUMO offset between PM6 and Y12 is relatively large (>0.5 eV), ensuring that electron transfer from donor to acceptor is not significantly impacted by the energetic changes during aging. However, charge generation in low-bandgap NFA systems critically relies on hole transfer from the acceptor to the donor, which is driven by the HOMO offset. In fresh blends without 1-CN, the HOMO offset is approximately 200 meV, providing adequate driving force for hole transfer. With aging, the 200 meV downward shift in the PM6 HOMO level reduces the HOMO offset to nearly zero, thereby eliminating the thermodynamic driving force for hole transfer from Y12 to PM6 and severely suppressing free charge generation efficiency. Previous studies have established that PM6-based non-fullerene OPVs require a sufficient HOMO offset to provide adequate driving force for efficient exciton dissociation at the donor-acceptor interface, and decreasing the offset reduces free charge generation efficiency and accelerates carrier recombination[49].

In contrast to PM6:Y12 blends, PM6:Y12 blends with 1-CN exhibit remarkable energetic stability throughout photodegradation. Fresh PM6:Y12 blends with 1-CN show HOMO levels at −5.3 eV for PM6 and −5.4 eV for Y12, similar to blends without 1-CN (Figure 2d). However, unlike PM6:Y12 blend, where the PM6 HOMO level shifts substantially upon aging, the active layer with 1-CN maintains energetic stability throughout the aging, preserving the HOMO levels of PM6. The 1-CN additive ensures that the thermodynamic driving force for hole transfer from acceptor to donor remains adequate throughout device operation, thereby sustaining efficient charge generation and extraction. The fast and efficient hole transfer process from Y12 to PM6 is hindered by the reduced HOMO offset in aged PM6:Y12 blends without 1-CN, whereas the 1-CN additive overcomes this limitation by maintaining the threshold HOMO offset necessary for efficient charge generation. Importantly, in both systems (with and without 1-CN), the acceptor preserves its energetics, further indicating that the primary performance-limiting degradation component is not the non-fullerene acceptor but rather the polymer donor PM6, and the addition of a 1-CN can uniquely stabilize the PM6 energetic alignment, presumably by improving morphology.

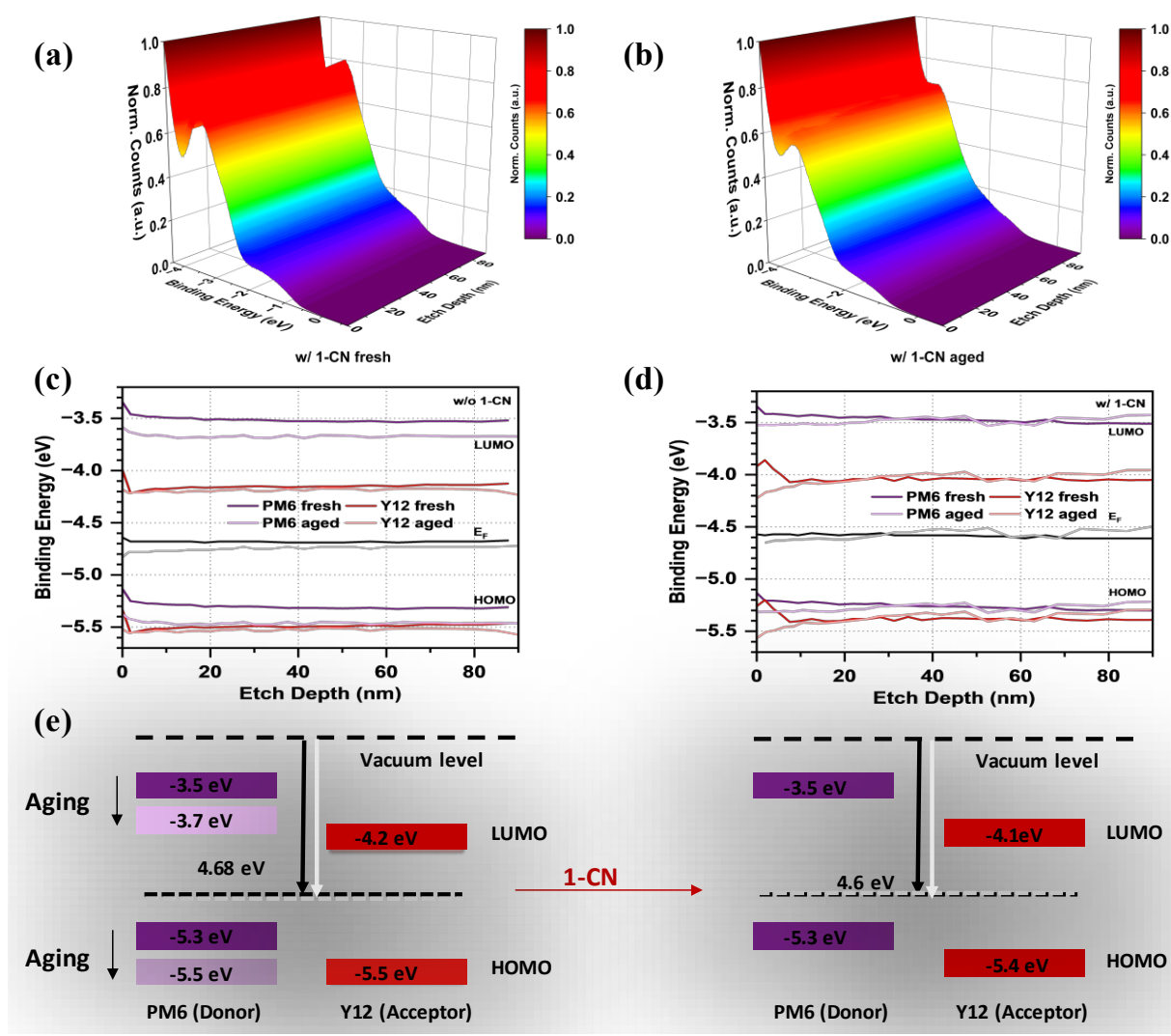

**Figure 2.** Three-dimensional energetic map of (a) fresh and (b) aged PM6:Y12 with 1-CN. Energetic landscape of blend films (c) without 1-CN and (d) with 1-CN, from UPS depth Profile. (e) Graphical representation of the evolution of the energetic landscape with and without 1-CN upon aging. Dark colors represent fresh blends, whereas light colors show the parameters of the aged blends.

To further confirm that the observed energetic changes in the blend originate from the donor material, we performed UPS depth profiling on neat PM6 and Y12 films (Figure S5). Neat Y12 films remain energetically stable upon photoaging, preserving their HOMO level at −5.5 eV, further confirming that Y12 is not susceptible to the energetic degradation observed in the blends. However, neat PM6 films exhibit a HOMO level shift of approximately 200 meV upon photoaging, similar to their change in the PM6:Y12 blend films. It further indicates PM6 energetics are inherently unstable upon photoaging in both pristine and blend films that the degradation mechanism is intrinsic to the polymer donor material rather than being induced by interactions with the acceptor.

X-ray photoemission spectroscopy (XPS) was performed to probe potential changes in the chemical composition of the active layer films upon photodegradation and to assess whether chemical degradation pathways are significant contributors to device degradation. Elemental composition analysis via XPS reveals that the C 1s, S 2p, N 1s, and F 1s core level spectra remain essentially unchanged in fresh and aged PM6:Y12 blends, both prepared with and without 1-CN (Figure S6, S7). But O 1s spectra show significant broadening and increased intensity after photoaging. However, the absence of significant new chemical species or shifts in binding energy suggests that large-scale photochemical degradation of the active layer (such as extensive oxidation, halogenation, or decomposition reactions) is not the primary degradation pathway in PM6:Y12 blends under photoaging employed in this study. Instead, the observed changes in energetics are more likely associated with nanostructure (morphological degradation, phase separation changes and crystallinity loss).

To investigate the vertical compositional gradient of PM6 and Y12 across the active layer thickness, UPS depth profiling was performed. Previous studies have demonstrated that the UPS depth profile outperforms the XPS depth profile in determining the compositional profile. The high surface sensitivity of UPS (a probing depth of 1-2 nm compared to 5-10 nm for XPS) improves depth resolution and reduces the influence of adventitious carbon contamination. Additionally, for binary blends such as PM6:Y12, XPS depth profiling cannot fully differentiate the donor and acceptor, as both materials contain similar elements (C, S, F, O) with similar chemical environments, with only the nitrogen content in the acceptor providing a distinguishing feature. In contrast, UPS depth profiling exploits the electronic structure as the distinguishing feature for compositional quantification. Since the electronic structures of polymer donor and non-fullerene acceptor differ substantially in their valence band features, we utilize these distinct electronic signatures to deduce the relative volume fractions of donor and acceptor as a function of depth. Figure S8 shows the material percentage of donor and acceptor throughout the BHJ thickness. In both blends, with and without 1-CN, a PM6-rich surface layer forms, likely due to the donor polymer's lower surface energy relative to the acceptor [54]. Beyond this surface enrichment, the bulk of the active layer exhibits relatively homogeneous mixing of donor and acceptor with compositions close to the donor: acceptor ratio (1:1.2), confirming that the BHJ structure is maintained throughout most of the film thickness. Importantly, the compositional profiles remain qualitatively similar between fresh and aged films, supporting the conclusion that major compositional changes are not the dominant degradation mechanism.

## 2.3 Impact on the Active Layer Microstructure

We then employed grazing-incidence wide-angle X-ray scattering (GIWAXS) measurements to study the structural morphology of donor, acceptor, and the blends with and without an additive. The direct comparison of the blends prior to aging shows that the addition of CN significantly enhances long-range order during processing, thereby affecting aging. GIWAXS revealed that neat PM6 exhibits a strong loss of ordering upon degradation, as evidenced by a substantial decrease in scattering intensity, indicating a less ordered nanostructure (Figures 3e and 3i). To quantify the nanostructural degradation of PM6, we fitted the PM6 (100) lamellar stacking peak in the vertical scattering direction. In neat PM6, the peak amplitude decreases dramatically from $114.8 \pm 1.1$ a.u. (fresh) to $17.8 \pm 1.1$ a.u. (aged), corresponding to an 85% reduction, confirming a severe loss of lamellar order upon photoaging. In contrast, the peak positions and intensities in neat Y12 remain essentially identical between fresh and aged films, demonstrating that the Y12 nanostructure is unaffected by the applied aging protocol (Figures 3f and 3j).

In the PM6:Y12 blend without 1-CN, the amplitude of the PM6 lamellar peak in the vertical direction drops from $19.6 \pm 2.8$ a.u. (fresh) to $7.5 \pm 1.7$ a.u. (aged), a 62% reduction. In blends processed with 1-CN, the initial PM6 (100) amplitude is substantially higher ($40.9 \pm 1.4$) a.u., reflecting the enhanced molecular ordering induced by the additive. The molecular long-range ordering of Y12 is also significantly enhanced, as evidenced by multiple additional Y12 scattering features in Figure 3d compared to 3c. Upon aging, the PM6 (100) amplitude decreases only to $29.6 \pm 1.4$ a.u., corresponding to a much smaller 28% reduction. The Y12 nanostructure remains almost unchanged. This quantitative analysis thus confirms that 1-CN not only promotes stronger initial PM6 and Y12 ordering but also significantly mitigates PM6 nanostructural degradation during photoaging, preserving nanostructural order and complementing the retained energetics in the active layer. The results suggest that decreased nanostructural ordering of the PM6:Y12 blends upon aging disrupts the energetics, whereas retained morphology upon aging with an additive helps maintain the energetic order of the active layer.

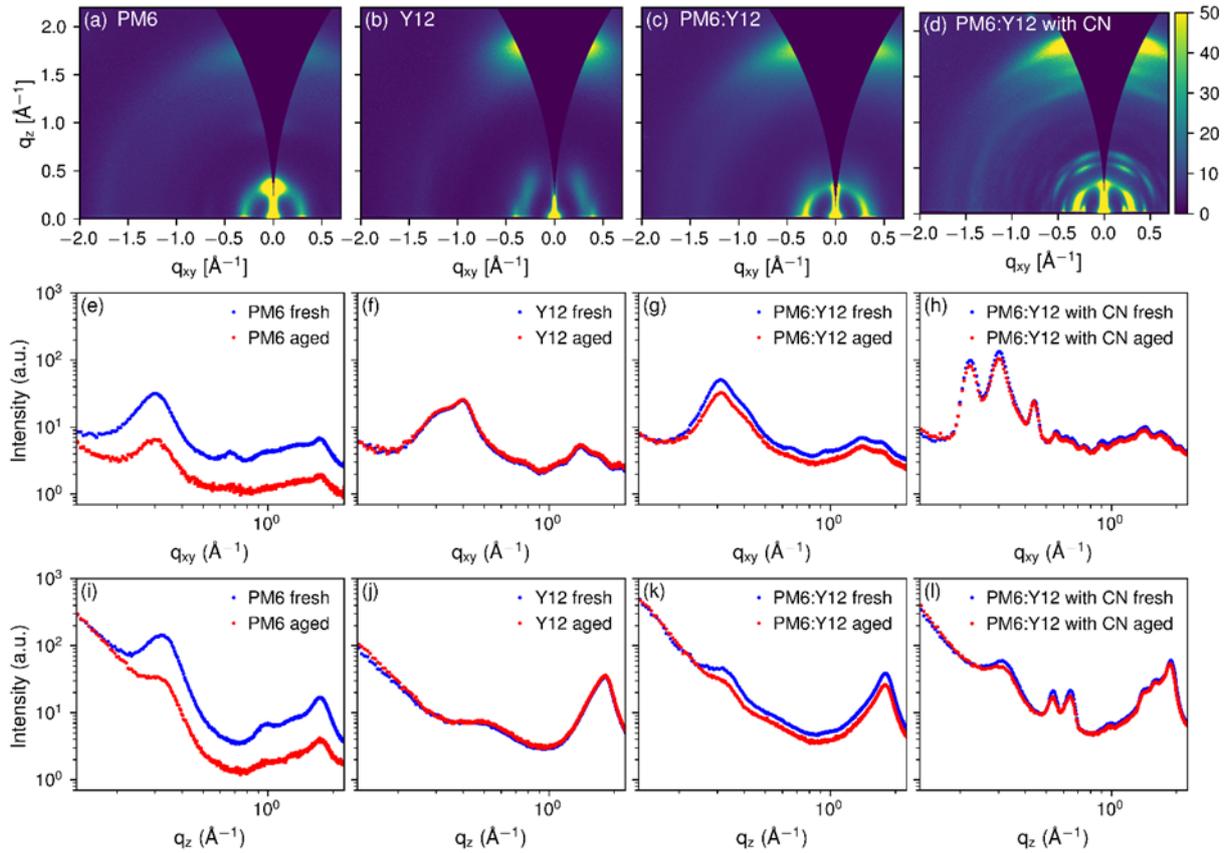

**Figure 3.** 2D GIWAXS data of fresh (a) PM6, (b) Y12, (c) PM6:Y12 without 1-CN, and (d) PM6:Y12 with 1-CN (comparison to aged 2D GIWAXS data in Figure S9). (e)-(h) Horizontal ($q_{xy}$) and (i)-(l) vertical ($q_z$) cake cuts comparing the fresh with the aged films.

## 2.4 Transient Absorption Studies

To link changes in film morphology to energetics, we employ femtosecond transient absorption spectroscopy. We photoexcite the sample at 400 nm, simultaneously probing the visible and NIR spectral regions. 2D Transient absorption maps for fresh and degraded PM6:Y12—both with and without the 1-CN are given in Figure 4a-d. We assign the negative photobleaching features at approximately 610 nm and 810 nm to the ground-state bleach of the PM6 and Y12 components, respectively. The broad photo-induced absorption (PIA) feature at 900-1050 nm results from overlapping exciton and free-carrier bands. We additionally observe a PIA band at 650-700 nm, which previous studies on Y-series NFAs have attributed to free carriers. We note, however, that the presence of strongly overlapping ground-state bleach (GSB) bands from both PM6 and Y12 substantially complicates the dynamics accessible from these bands; for this reason, we focus primarily on the dynamics of the 900-1050 nm PIA feature.

The kinetics of the 900-1050 nm PIA features are given in Figures 4e and 4f. The pristine blend films, both with and without 1-CN, exhibit clear biphasic dynamics, with a long-lived decay component (100-1000 ps) apparent in both systems. We attribute this long-lived component to the formation of free charges within the film, with the early-time component to excitonic recombination. For both blends, photoaging quenches this long-lived component, indicating inhibited free-charge generation. This loss of long-lived charge carrier signal directly correlates with the energetic changes observed in UPS depth profiling: the reduction in HOMO offset eliminates the driving force for hole transfer, leading to increased geminate recombination and reduced free carrier yield. Importantly, this difference in kinetics upon aging is substantially less pronounced for PM6:Y12 blends processed with 1-CN, further supporting the conclusion that the addition of 1-CN improves the film's robustness against environmental degradation by preserving both the energetic driving force for charge separation and the nanostructural stability necessary for efficient charge generation and extraction.

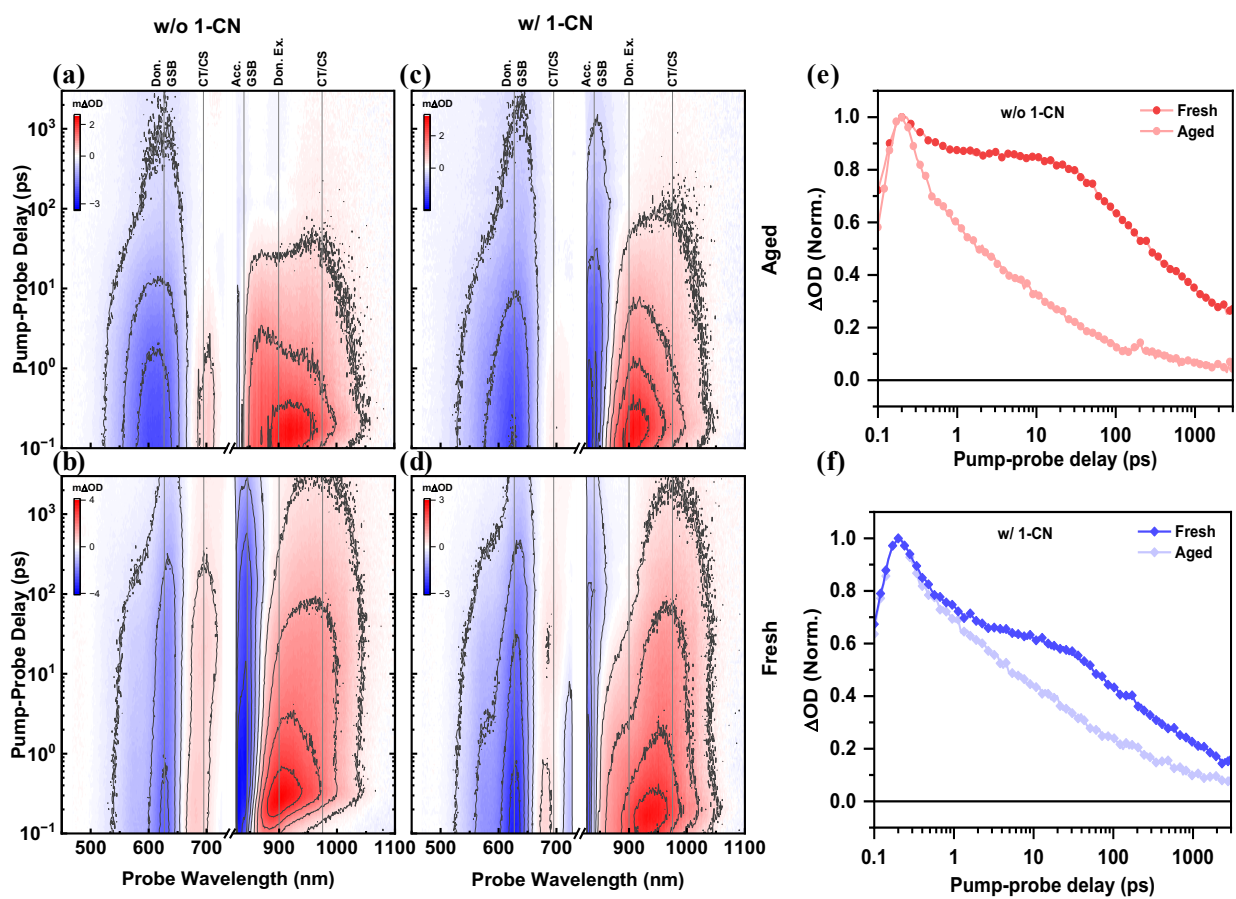

**Figure 4.** (a)-(d) 2D transient absorption maps of fresh and aged PM6:Y12 without 1-CN and with 1-CN. TA kinetics of the photo absorption feature at 950 nm for (e) without 1-CN (f) with 1-CN.

## 2.5 Device Performance

To correlate the structural and energetic changes with device performance, we fabricated conventional organic photovoltaic devices with the structure ITO/PEDOT:PSS/PM6:Y12/PDINN/Ag, as shown in Figure 5a. Figure 5b represents the illuminated current density-voltage (J-V) characteristics of fresh and aged devices. Photovoltaic parameters of fresh and aged devices (with and without 1-CN) are shown in Figure 5c-f. The fresh PM6:Y12 devices exhibit a maximum PCE of 14.25%, along with a $J_{SC}$ of -27.14 mA cm$^{-2}$, a $V_{OC}$ of 0.84 V, and an FF of 64.48%. While the addition of 1-CN has improved efficiency in pristine devices, achieving a maximum PCE of 17.3%, accompanied by a $J_{SC}$ of -28.17 mA cm$^{-2}$, a $V_{OC}$ of 0.84 V, and an FF of 72.74%. After 15 hours of illumination in ambient air, devices fabricated without 1-CN display a pronounced reduction in all J-V parameters, indicating severe photodegradation. However, the 1-CN additive mitigates the significant decrease in $J_{SC}$ and FF, thereby improving device performance. It is noteworthy that incorporating the 1-CN additive not only enhances device performance but also significantly improves long-term stability compared to additive-free devices. Detailed stability analyses conducted on devices with and without 1-CN (Figure S10) demonstrate that devices with 1-CN retain a higher percentage of their initial PCE throughout the aging period than devices without 1-CN. The external quantum efficiency spectrum of the fresh and aged devices is depicted in Figure S11 and shows a significant decrease upon aging, which matches well with the decrease in the $J_{SC}$ obtained from J-V characteristics.

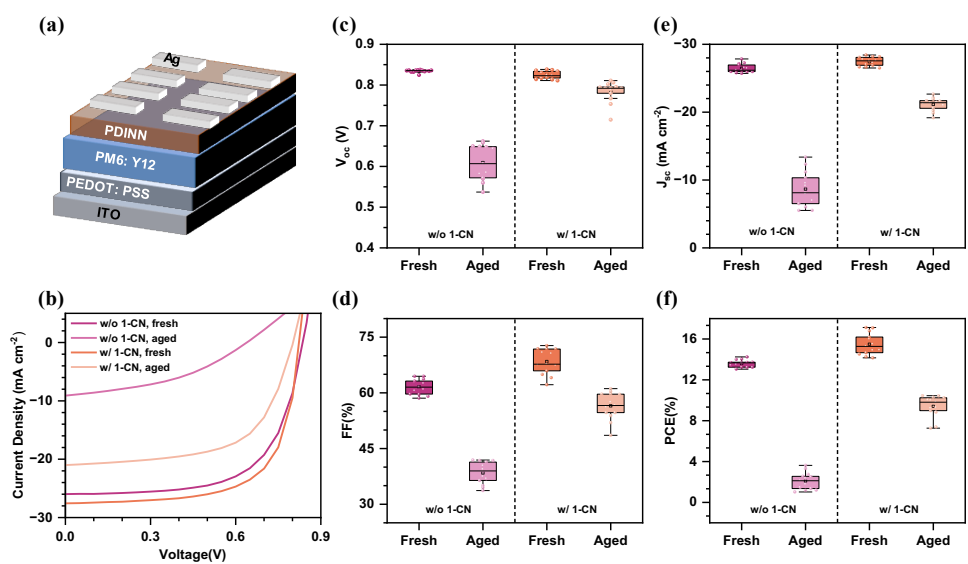

**Figure 5.** (a) A schematic of the device architecture. (b) current density–voltage (J-V) characteristics of fresh and aged devices (with and without 1-CN). (c-f) Solar cell parameters of fresh and aged devices (with and without 1-CN). The statistics are based on 13 individual cells.

## 3. Conclusion

This study reveals that the use of 1-chloronaphthalene as an additive significantly influences the photostability of PM6:Y12 bulk heterojunction organic solar cells, particularly the evolution of the energetic landscape during photodegradation. Our UPS depth-profiling measurements reveal the realistic energetics of the donor and acceptor in the bulk heterojunction and show that photodegradation in PM6:Y12 blends is donor-limited, with the polymer PM6 undergoing a critical HOMO deepening upon photoaging under ambient conditions, while the non-fullerene acceptor Y12 maintains remarkable energetic stability. This energetic misalignment eliminates the thermodynamic driving force for hole extraction, thereby reducing device performance. Interestingly, the addition of 1-CN preserves the HOMO levels of PM6 throughout photoaging, maintaining the threshold HOMO offset necessary for efficient hole transfer from Y12 to PM6 and thereby sustaining free charge generation efficiency. GIWAXS revealed that CN enhances nanostructural order during processing. Neat PM6 exhibits stronger nanostructural degradation with disordered molecular packing, whereas the nanostructure of neat Y12 remains unchanged upon aging. The PM6:Y12 blends exhibit greater nanostructural degradation, whereas the 1-CN additive significantly mitigates this degradation, preserving ordered Y12 domains and maintaining PM6 stacking integrity during aging. These studies were complemented by transient absorption spectroscopy measurements, which showed reduced charge-generation efficiency in aged PM6:Y12 blends but better charge generation and hole transfer in blends with 1-CN. The results suggest that future molecular design efforts should focus on developing donors with enhanced energetic stability under operational conditions, and that processing strategies should be optimized not only for initial performance but also for long-term stability.

## 4. Experimental Section

*Materials*

Organic photovoltaic materials PM6 and Y12 were purchased from Brilliant Maters. PDINN was purchased from Solarmer Materials Inc. PEDOT:PSS (Clevios PVP Al4083) was purchased from Heraeus. Additive 1-CN, Chloroform, and methanol were purchased from Sigma Aldrich Inc. All materials and solvents were used as received without further purification.

Full Material names:

PM6: poly[(2,6-(4,8-bis(5-(2-ethylhexyl-3-fluoro)thiophen-2-yl)-benzo[1,2-b:4,5-b']dithiophene))-alt-(5,5-(1',3'-di-2-thienyl-5',7'-bis(2-ethylhexyl)benzo[1',2'-c:4',5'-c']dithiophene-4,8-dione)]

Y12: 2,2'-((2Z,2'Z)-((12,13-bis(2-butyloctyl)-3,9-diundecyl-12,13-dihydro-[1,2,5]thiadiazolo[3,4-e]thieno[2",3":4',5']thieno[2',3':4,5]pyrrolo[3,2-g]thieno[2',3':4,5]thieno[3,2-b]indole-2,10-diyl)bis(methanylylidene))bis(5,6-difluoro-3-oxo-2,3-dihydro-1H-indene-2,1-diylidene))dimalononitrile

PDINN: N,N'-Bis{3-[3-(Dimethylamino)propylamino]propyl}perylene-3,4,9,10-tetracarboxylic diimide

PEDOT:PSS: poly(3,4-ethylenedioxythiophene):poly(styrenesulfonate)

*Device Fabrication*

Organic solar cells were fabricated on pre-patterned indium-doped tin oxide (ITO)- coated glass substrates with the conventional structure: ITO/PEDOT:PSS/PM6:Y12/PDINN/Ag. Before fabrication, the substrates were cleaned in an ultrasonic bath with Distilled water, acetone, and isopropanol for 10 min each, and then treated with ultraviolet (UV) ozone for 10 min. PEDOT: PSS aqueous solution was filtered through a 0.22 μm polytetrafluoroethylene (PTFE) filter and spin-coated onto precleaned ITOs. The films were annealed at 150°C for 20 minutes. The substrates were then transferred into a nitrogen-filled glovebox. The active layers were spin-coated from a PM6:Y12 (1:1.2 wt%, 16 mg mL$^{-1}$) chloroform solution at 3000 rpm for 30 s. For additive engineering, 0.5% v/v 1-CN is also added to the solution. The active layers were thermally annealed at 100°C for 10 minutes. Then, the PDINN film was deposited by spin-coating a solution of 1 mg mL$^{-1}$ PDINN in methanol. Finally, the silver electrode was deposited by thermal evaporation. The effective device area was 0.045 cm$^2$, defined by the overlapping area of ITO and Ag.

*Characterization Methods*

*Current density–voltage (J-V) measurements*: J-V measurements were carried out in an ambient condition under simulated AM 1.5 light with an intensity of 100 mW cm$^{-2}$ (ABET TECHNOLOGIES Sun 3000 AAA solar simulator). The intensity was calibrated using a Si reference cell (NIST traceable, VLSI) and corrected by determining the spectral mismatch between the solar spectrum, reference cell, and spectral response of the device. The external quantum efficiency spectra were recorded using monochromatic light from a halogen lamp from 400 to 1100 nm; the reference spectra were calibrated using a NIST-traceable Si diode (Thorlabs).

*Ultraviolet-visible (UV-Vis) absorption spectroscopy*: optical absorption spectra of neat and blend films were obtained using a Shimadzu UV-3101 PC spectrometer.

*UPS/XPS measurements*: The films were transferred to an ultrahigh vacuum chamber (ESCALAB 250Xi by Thermo Scientific) with a base pressure of 2 × 10$^{-10}$ mbar for XPS and UPS measurements. XPS measurements were carried out using an XR6 monochromated Al Kα source (hv = 1486.6 eV) with a spot size of 650 μm. Energy levels of 100 and 20 eV were used for the survey and core level spectra, respectively. UPS measurements were performed using a double-differentially pumped He discharge lamp (h$v$ = 21.22 eV) with a pass energy of 2 eV and a bias of −5 V, thereby ensuring detection of the secondary-electron cut-off.

*UPS/XPS Depth Profiling*: UPS and XPS depth profiling were performed using the MAGCIS dual-mode ion source (ESCALAB 250Xi, Thermo Scientific) operated in argon gas cluster ion beam mode, which has been shown to minimize etching-induced material damage. The cluster mode was used at an energy of 4 keV with clusters (in the order of Ar$_{2000}$) and an etch crater size of (2 × 2) mm². To create a depth profile, XPS or UPS measurements were altered with cluster etch steps.

*GIWAXS*: Grazing incidence wide-angle X-ray scattering (GIWAXS) was performed on a laboratory system at the University of Bayreuth (Xeuss 3.0, Xenocs SAS, Grenoble, France) with a GeniX3D HFVL Cu Kα source (λ=1.54 Å), a Dectris EIGER 2R 1M detector, and a sample-to-detector distance of 72 mm and a beam size of 500 μm. Scattering experiments were carried out at room temperature under vacuum on samples on silicon substrates with a length of 5 mm in beam direction. The incident angle was set to 0.18° (above the critical angle of about 0.16°), which probes the full depth of the films. The presented q-profiles are cake cuts covering an azimuthal angle of 70–110° for the cuts in the vertical direction and 0–20° as well as 160–180° for the cuts in the horizontal direction.

The fits of the (100) lamellar PM6 peak are Pseudo-Voigt fits, described by the following expression for the single peak:

$$f(q) = A \cdot [\eta \cdot L(q) + (1 - \eta) \cdot G(q)] \text{ with } 0 < \eta < 1$$

$$G(q) = exp\left[-ln(2) \cdot \left(\frac{q-c}{b}\right)^2\right], \quad L(q) = \frac{1}{1 + \left(\frac{q-c}{b}\right)^2}$$

Where $A$ is the peak amplitude, $c$ is the peak position, $2b$ is the full width at half maximum of the Pseudo-Voigt peak, and $\eta$ is the Pseudo-Voigt mixing parameter. We fitted one Pseudo-

Voigt peak with an additional background consisting of a $q^{-5}$, an additional $q^{-2}$ decay and an offset.